\documentclass[twoside,12pt,leqno]{article}
\advance\topmargin by -\headheight\advance\topmargin by -\headsep

\usepackage{amsmath}
\usepackage{amsthm}
\usepackage{amssymb}
\usepackage{amscd}
\usepackage{graphicx}
\usepackage{epsfig}

\numberwithin{equation}{section}
\allowdisplaybreaks

\newtheorem{theorem}{Theorem}[section]

\theoremstyle{definition}

\newtheorem{remark}{Remark}[section]
\newtheorem*{remarku}{Remark}
\begin{document}
\newcommand{\be}{\begin{eqnarray}}
\newcommand{\en}{\end{eqnarray}\vs 0.5 cm}
\newcommand{\non}{\nonumber}
\newcommand{\no}{\noindent}
\newcommand{\vs}{\vskip}
\newcommand{\hs}{\hspace}
\newcommand{\e}{\'{e}}
\newcommand{\ef}{\`{e}}
\newcommand{\p}{\partial}
\newcommand{\un}{\underline}
\newcommand{\NR}{{\mathbb R}}
\newcommand{\NA}{{{\dl A}}}
\newcommand{\NP}{{{\dl P}}}
\newcommand{\NC}{{{\C}}}
\newcommand{\NT}{{{\dl T}}}
\newcommand{\NZ}{{{\dl Z}}}
\newcommand{\NH}{{{\dl H}}}
\newcommand{\NN}{{{\dl N}}}
\newcommand{\NS}{{{\dl S}}}
\newcommand{\NW}{{{\dl W}}}
\newcommand{\NV}{{{\dl V}}}
\newcommand{\qq}{\begin{eqnarray}}
\newcommand{\de}{\bar\partial}
\newcommand{\da}{\partial}
\newcommand{\ee}{{\rm e}}
\newcommand{\Ker}{{\rm Ker}}
\newcommand{\qqq}{\end{eqnarray}}
\newcommand{\ssigma}{\mbox{\boldmath $\sigma$}}
\newcommand{\llambda}{\mbox{\boldmath $\lambda$}}
\newcommand{\aalpha}{\mbox{\boldmath $\alpha$}}
\newcommand{\xx}{\mbox{\boldmath $x$}}
\newcommand{\xxi}{\mbox{\boldmath $\xi$}}
\newcommand{\kk}{\mbox{\boldmath $k$}}
\newcommand{\tr}{\hbox{tr}}
\newcommand{\ad}{\hbox{ad}}
\newcommand{\Lie}{\hbox{Lie}}
\newcommand{\w}{{\rm w}}
\newcommand{\CA}{{\cal A}}
\newcommand{\CB}{{\cal B}}
\newcommand{\CC}{{\cal C}}
\newcommand{\CE}{{\cal E}}
\newcommand{\CF}{{\cal F}}
\newcommand{\CG}{{\cal G}}
\newcommand{\CH}{{\cal H}}
\newcommand{\CI}{{\cal I}}
\newcommand{\CJ}{{\cal J}}
\newcommand{\CK}{{\cal K}}
\newcommand{\CL}{{\cal L}}
\newcommand{\CM}{{\cal M}}
\newcommand{\CN}{{\cal N}}
\newcommand{\CO}{{\cal O}}
\newcommand{\CP}{{\cal P}}
\newcommand{\CQ}{{\cal Q}}
\newcommand{\CR}{{\cal R}}
\newcommand{\CS}{{\cal S}}
\newcommand{\CT}{{\cal T}}
\newcommand{\CU}{{\cal U}}
\newcommand{\CV}{{\cal V}}
\newcommand{\CW}{{\cal W}}
\newcommand{\CX}{{\cal X}}
\newcommand{\CY}{{\cal Y}}
\newcommand{\CZ}{{\cal Z}}
\newcommand{\s}{\hspace{0.05cm}}
\newcommand{\m}{\hspace{0.025cm}}
\newcommand{\ch}{{\rm ch}}
\newcommand{\ra}{{\rightarrow}}
\newcommand{\mt}{{\mapsto}}
\newcommand{\hf}{{_1\over^2}}
\newcommand{\Di}{{\slash\hs{-0.21cm}\partial}}
\newcommand{\bx}{{\bf x}}
\newcommand{\parr}{\partial_{r}\, } 
\newcommand{\parl}{\partial_{l}\, } 
\newcommand{\AL}{A_{_l}} 
\newcommand{\AR}{A_{_r}} 
\newcommand{\Ad}{{\rm Ad}} 
\newcommand{\gp}{{g_{_+}}} 
\newcommand{\gm}{{g_{_-}}} 
\newcommand{\gpinv}{{g_{_{+}}^{-1}}}
\newcommand{\gminv}{{g_{_{-}}^{-1}}} 
\newcommand{\ggl}{{\gl g}} 
\newcommand{\jlp}{{\gpinv\parl\gp}} 
\newcommand{\jrp}{{\gpinv\parr\gp}}
\newcommand{\jlm}{{\gminv\parl\gm}} 
\newcommand{\jrm}{{\gminv\parr\gm}}

\setcounter{page}{1}
\thispagestyle{empty}



\markboth{I. Calvo, F. Falceto}{Poisson-Dirac branes in Poisson-Sigma models}
\label{firstpage}
$ $
\bigskip

\bigskip

\centerline{\Large Poisson-Dirac branes}
\bigskip
\centerline{\Large in Poisson-Sigma models}

\bigskip
\bigskip
\centerline{\large by}
\bigskip
\bigskip
\centerline{\large Iv\'an Calvo and  Fernando Falceto}
\centerline{Depto. F\'{\i}sica Te\'orica, Univ. Zaragoza,}
\centerline{E-50009 Zaragoza, Spain}
\centerline{E-mail: {\tt icalvo@unizar.es, falceto@unizar.es}}

\vspace*{.7cm}

\begin{abstract}
\noindent
We analyse the general boundary conditions (branes) consistent with
the Poisson-sigma model and study the structure of the phase space of
the model defined on the strip with these boundary
conditions. Finally, we discuss the perturbative quantization of the
model on the disc with a Poisson-Dirac brane and relate it to
Kontsevich's formula for the deformation quantization of the Dirac bracket induced on the brane.
\end{abstract}

\pagestyle{myheadings}
\section{Introduction}

Poisson-Sigma models (\cite{Ikeda}, \cite{Strobl}) are topological
field theories based on a bundle map from the tangent bundle of a
surface $\Sigma$ to the cotangent bundle of a Poisson manifold $M$. A
number of two-dimensional gauge theories such as pure gravity, WZW
$G/G$ (locally) and Yang-Mills (up to addition of a non-topological
term containing the Casimir of the Poisson structure) are particular
cases of Poisson-Sigma models.

The mathematical interest of the Poisson-Sigma model resides in the
fact that it naturally encodes the Poisson Geometry of the target and
allows to unravel it by means of techniques from Classical and Quantum
Field Theory. This feature has allowed to shed considerable light on
previous mathematical results, such as the interpretation of
Kontevich's star product in terms of Feynman diagrams by Cattaneo and
Felder in \cite{CaFe}. Moreover, the same authors showed in
\cite{CaFe2} the connection between the structure of the phase space
of the model and the symplectic groupoid (when the latter exists)
integrating the target Poisson manifold. Their ideas inspired to a
large extent the crucial work \cite{CrFe2} of Crainic and Fernandes on
the integrability of Lie algebroids.

The exposition below (based on \cite{CaFa}) is concerned with the
boundary conditions (BC) allowed in the Poisson-Sigma model. We
generalize the results of Cattaneo and Felder (\cite{CaFe3}) and show
that the base map can be consistently restricted at the boundary to
almost arbitrary submanifolds (branes) of the target. A thorough
analysis of the model with general BC is provided in the Lagrangian as
well as in the Hamiltonian formalism. It turns out that the structure
of the phase space is related to the Poisson bracket on the brane
obtained by reduction rather than to the original Poisson bracket on
$M$.

In ref. \cite{CaFe3} the claim was that only coisotropic branes were
admissible. In those cases the reduced Poisson bracket on the brane is
just the original Poisson bracket restricted to the subset of gauge
invariant functions.

Our general setup allows in particular the other extreme situation,
i.e. no gauge transformations at all. We will use the term 
{\it Poisson-Dirac} branes following \cite{CrFe}. This
case leads to a nontrivial reduction of the Poisson bracket of $M$ to
the brane (Dirac bracket).

In these pages we want to stress the interest of this class of branes
for the quantization of the model. We devote the last section to the
discussion on the perturbative quantization on the disc with a
Poisson-Dirac brane and conjecture that, after a suitable choice of
the unperturbed part of the action, it yields the Kontsevich's star
product for the Dirac bracket induced on the brane.

\section{Poisson-Sigma models} \label{PSmodels}
 
The Poisson-Sigma model is a two-dimensional topological Sigma model
defined on a surface $\Sigma$ and with a finite dimensional Poisson
manifold $(M,\Gamma)$ as target.

The fields of the model are given by a bundle map $(X,\psi): T\Sigma
\rightarrow T^{*}M$ consisting of a base map $X:\Sigma \rightarrow M$
and a 1-form $\psi$ on $\Sigma$ with values in the pullback by $X$ of
the cotangent bundle of $M$. If $X^{i}$ are local coordinates in $M$,
$\sigma ^{\mu},\ \mu=1,2$ local coordinates in $\Sigma$, $\Gamma^{ij}$
the components of the Poisson structure in these coordinates and
$\psi_{i}=\psi_{i\mu}d{\sigma}^{\mu}$, the action reads
\begin{eqnarray} \label{PScoor}
S_{P\sigma}(X,\psi)=\int_\Sigma dX^{i}\wedge \psi_{i}-
{1\over2}\Gamma^{ij}(X)\psi_{i}\wedge \psi_{j}
\end{eqnarray}

It is straightforward to work out the equations of motion in the bulk:
\begin{eqnarray}
&&dX^{i}+\Gamma^{ij}(X)\psi_{j}=0 \label{eom}\cr 
&&d\psi_{i}+{1\over2}\partial_{i}\Gamma^{jk}(X)\psi_{j}\wedge \psi_{k}=0
\end{eqnarray}

One can show (\cite{BojoStrobl}) that for solutions
of (\ref{eom}) the image of $X$ lies within 
one of the symplectic leaves  of the foliation of $M$.

Under the infinitesimal transformation
\begin{eqnarray} \label{symmetry}
&&\delta_{\epsilon}X^{i}=
\Gamma^{ji}(X)\epsilon_{j}\cr
&&\delta_{\epsilon}\psi_{i}=d\epsilon_{i}+\partial_{i}\Gamma^{jk}(X)\psi_{j}
\epsilon_{k}
\end{eqnarray}
where $\epsilon=\epsilon_{i}dX^i$ is a section of $X^*(T^*M)$,
the action (\ref{PScoor}) transforms by a boundary term
\begin{eqnarray} \label{symmS}
\delta_{\epsilon}S_{P\sigma}=-\int_\Sigma d(dX^i \epsilon_i).
\end{eqnarray}

\section{Boundary conditions} \label{BC}

In surfaces with boundary (in this section we restrict ourselves to a
boundary with one connected component) a new term appears in the
variation of the action under a change of $X$ when performing the
integration by parts:
\qq
\delta_{X} S=\int_{\partial\Sigma}\delta X^i\psi_i-
\int_\Sigma\delta X^i(d\psi_{i}+
{1\over2}\partial_{i}\Gamma^{jk}(X)\psi_{j}\wedge \psi_{k})
\qqq
 
Let us take the field
$X|_{\partial \Sigma}:\partial \Sigma\rightarrow C$
for an arbitrary (for the moment) closed embedded submanifold $C$ of
$M$ (brane, in a more stringy language). Then $\delta X\in T_{_X} C$
at every point of the boundary and, if the surface term is to vanish,
the contraction of $\psi=\psi_{i}dX^i$ with vectors tangent to the
boundary (that we will denote by $\psi_t=\psi_{it}dX^i$) must belong
to $N^*_{_X}C$. Here $N^*C$ is the conormal bundle of $C$, i.e. the
subbundle of $T^{*}_{C}M$ whose fibers are the covectors that kill all
vectors tangent to $C$.

On the other hand, by continuity, the equations of motion in the bulk
must be satisfied also at the boundary. In particular, $\partial_t
X=\Gamma^\sharp_{_X}\psi_t$, where by $\partial_t$ we denote the
derivative along the vector on $\Sigma$ tangent to the boundary and
$\Gamma^\sharp:T^{*}M \rightarrow TM$ is given by the contraction with
the first factor of $\Gamma$. As $\partial_t X$ belongs to $T_{_X}C$
it follows that $\psi_t\in\Gamma^{\sharp-1}_{_X}(T_{_X}C)$.

Both conditions for $\psi_t$ imply that
\begin{eqnarray}\label{BCpsi}
\psi_t(m)\in\Gamma^{\sharp-1}_{_{X(m)}}(T^{}_{_{X(m)}}C)\cap N_{_{X(m)}}^*C,
\mbox{ for any }m\in \partial\Sigma 
\end{eqnarray}
which is the boundary condition we shall take for $\psi_t$.

Now, we must find out how these BC restrict the gauge transformations
(\ref{symmetry}) at the boundary. In order to cancel the boundary term
(\ref{symmS}) $\epsilon\vert_{\partial\Sigma}$ must be a smooth
section of $N^{*}C$ and if (\ref{symmetry}) is to preserve the
boundary condition of $X$, $\epsilon\vert_{\partial\Sigma}$ must
belong to $\Gamma^{\sharp-1}(TC)$. Hence,
\begin{eqnarray}\label{BCepsilon}
\epsilon(m)=\Gamma^{\sharp-1}_{_{X(m)}}(T^{}_{_{X(m)}}C)\cap N_{_{X(m)}}^*C,
 \ \forall m \in \partial\Sigma. 
\end{eqnarray}

Now we have the following
\begin{theorem}\label{constrank}
If $${\rm dim}(\Gamma^\sharp_p(N_p^*C)+ T_pC)=k, \ \forall p \in C$$
then the gauge transformations satisfying (\ref{BCepsilon}) also
preserve the BC (\ref{BCpsi}) for $\psi$, i.e. we have consistent
branes for the Poisson-Sigma model.
\end{theorem}

\begin{proof}
The proof can be found in ref. \cite{CaFa}.
\end{proof}

Some particular cases that have been considered in the literature
are the free boundary conditions $C=M$ and the coisotropic one
$\Gamma^\sharp (N^*C)\subset TC$. Both fulfill the constant
dimension requirement of theorem \ref{constrank}.

As mentioned in the introduction, a novel kind of brane particularly
interesting in the context of quantization is obtained when $C$ is a
{\it constant rank Poisson-Dirac submanifold}:
$\Gamma^\sharp_p(N_p^*C)\cap T_pC=0 \quad {\rm and}\quad$
${\rm dim}(\Gamma^\sharp_p(N_p^*C) + T_pC)=k, \ \forall p \in C$.
In this case there is no gauge transformation acting on the brane.

Before going to the Hamiltonian analysis of the theory we will need
some results on the algebraic reduction of Poisson manifolds that will be
summarized in the next section.

\section{Reduction of Poisson manifolds.}

Let $C$ be a closed submanifold of $(M,\Gamma)$. We adopt the notation
$\CA = C^{\infty}(M)$ and take the ideal (with respect to the
pointwise product of functions in $\CA$), $\CI =\{f \in \CA \vert f(p)
= 0,\ \forall p \in C\}$.

Define $\CF\subset\CA$ as the
set of {\it first-class functions}, also called the {\it normalizer} of $\CI$,
$$\CF:=\{f\in\CA \vert \{f,\CI\}\subset\CI\}$$ $\CF$ is a Poisson
subalgebra of $\CA$ and $\CF\cap\CI$ is a Poisson ideal of
$\CF$. Then, we have canonically defined a Poisson bracket in the
quotient $\CF/(\CF\cap\CI)$. We would like to find a Poisson bracket
on $C^{\infty}(C)\cong {\cal A}/{\cal I}$ (or, at least, in a subset
of it). To that end we define an injective map \qq\label{mapphi}
 \begin{matrix}\phi:&\CF/(\CF\cap\CI)&\longrightarrow&{\cal A}/{\cal I}\cr
&f+\CF\cap\CI&\longmapsto&f+\CI\end{matrix}
\qqq
$\phi$ is an homomorpism of abelian, associative algebras with unit 
and then induces a Poisson algebra structure $\{.,.\}_C$ on the 
image that will be denoted by $\CC(\Gamma,M,C)$,
\qq\label{ourDirac}
\{ f_1+\CI, f_2+\CI\}_{_C}=
\{ f_1, f_2\}+\CI.\qquad f_1,f_2\in \CF.
\qqq

In some cases it is possible to give a simple characterization of the image of $\phi$ as shows the following

\begin{theorem}\label{observables}

If ${\rm dim}(\Gamma^\sharp_p(N_p^*C)+ T_pC)= k,\ \forall p\in C$,
then $$\phi ({\CF}/{\CF}\cap{\CI})=\{f + \CI\vert \{f,\CF \cap
\CI\}\subset \CI\}$$ In other words, $\CC(\Gamma,M,C)$ is the set of
gauge invariant functions on $C$.
\end{theorem}

\begin{proof}
See \cite{CaFa}.
\end{proof}

For the constant rank Poisson-Dirac submanifold described at the end
of the previous section there are not gauge transformations acting on
it and the map $\phi$ is onto.

\section{Hamiltonian analysis}

We proceed to the hamiltonian study of the model when $\Sigma =
[0,\pi]\times {\NR}$ (open string). The fields in the hamiltonian
formalism are a smooth map $X:[0,\pi]\rightarrow M$ and a 1-form
$\psi$ on $[0,\pi]$ with values in the pull-back $X^*(T^*M)$; in
coordinates, $\psi=\psi_{i\sigma} dX^i d\sigma$.

Consider the infinite dimensional manifold of smooth maps $(X,\psi)$
with canonical symplectic structure $\Omega$. The action of $\Omega$
on two vector fields (denoted for shortness $\delta,\delta'$) reads
\begin{eqnarray} \label{Omega}
\Omega(\delta,\delta') = \int_{0}^{\pi}(\delta X^i \delta' 
\psi_{i\sigma} - \delta' X^i \delta \psi_{i\sigma})d\sigma
\end{eqnarray}

The phase space $P(M;C_0,C_{\pi})$ of
the theory is defined by the constraint:
\begin{eqnarray} \label{constraint}
\partial_{\sigma}X^i + \Gamma^{ij}(X)\psi_{j\sigma} = 0
\end{eqnarray}
and BC as in section 3 with $X(0)\in C_0$ and $X(\pi)\in C_\pi$ for
two closed submanifolds $C_{u}\subset M$, ${u}=0,\pi$.

This geometry, with a boundary consisting of two connected components,
raises the question of the relation between the BC at both ends. Note
that due to eq. (\ref{constraint}) $X$ varies in $[0,\pi]$ inside a
symplectic leaf of $M$. This implies that in order to have solutions
the symplectic leaf must have non-empty intersection both with $C_0$
and $C_{\pi}$.  In other words, only points of $C_0$ and $C_{\pi}$
that belong to the same symplectic leaf lead to points of
$P(M;C_0,C_{\pi})$. In the following we will assume that this
condition is met for every point of $C_0$ and $C_\pi$ and
correspondingly for the tangent spaces. That is, if we define the maps
\begin{eqnarray}
J_0:P(M,C_0,C_\pi)&\longrightarrow & C_0\cr
(X,\psi)&\longmapsto& X(0).
\end{eqnarray}
and 
\begin{eqnarray}
J_\pi:P(M,C_0,C_\pi)&\longrightarrow & C_\pi\cr
(X,\psi)&\longmapsto& X(\pi).
\end{eqnarray}
we assume that both maps are surjective submersions.

The canonical symplectic 2-form is only presymplectic when restricted
to $P(M;C_0,C_{\pi})$. The kernel is given by:
\begin{eqnarray} \label{Hamsymm}
&&\delta_{\epsilon}X^{i}=\epsilon_{j}\Gamma^{ji}(X)\cr
&&\delta_{\epsilon}\psi_{i\sigma}=\partial_{\sigma}\epsilon_{i}+\partial_{i}\Gamma^{jk}(X)\psi_{j\sigma}\epsilon_{k}
\end{eqnarray}
where $\epsilon$, a section of $X^*(T^*M)$, is subject to the BC
$$\epsilon({u})\in \Gamma_{_{X({u})}}^{\sharp -1}(T^{}_{_{X({u})}}
C_{u}) \cap N^*_{_{X({u})}}C_{u},\ {\rm for}\ {u}=0,\pi$$ 
which is consistent with the results of section 3 in the lagrangian
formalism.

The presymplectic structure induces a Poisson algebra $\CP$ in a
subset of the functions on the phase space $P(M,C_0,C_\pi)$. On the
other hand, we have the Poisson algebras $\CC(\Gamma,M,C_0)$ and
$\CC(\Gamma,M,C_\pi)$. What is the relation among them?

We first analyse when a function $F(X,\psi)=(J_0^*f)(X,\psi)=f(X(0))$,
$f\in{C}^\infty(M)$ belongs to $\CP$, i.e. when it has a hamiltonian
vector field $\delta_F$.  Solving the corresponding equation we see
that the general solution is of the form (\ref{Hamsymm}) with
\begin{equation}\label{cond0}
\epsilon(0)-df_{_{X(0)}}\in
N^*_{_{X(0)}}C_0,
\qquad\epsilon(0)\in \Gamma_{_{X(0)}}^{\sharp -1}(T^{}_{_{X(0)}} C_0)
\end{equation}
and  
$$\epsilon(\pi)\in
\Gamma_{_{X(\pi)}}^{\sharp -1}(T^{}_{_{X(\pi)}} C_\pi)
\cap N^*_{_{X(\pi)}}C_\pi.$$

Assuming ${\rm dim}(\Gamma^\sharp(N^*_pC_0)+T_pC_0)=const.$, equation
(\ref{cond0}) can be solved in $\epsilon(0)$ if and only if $F$ is a
gauge invariant function (i.e. it is invariant under (\ref{Hamsymm})).
This is equivalent to saying that $f+\CI_0$ belongs to the Poisson
algebra $\CC(\Gamma,M,C_0)$.  (Here $\CI_0$ is the ideal of functions
that vanish on $C_0$).

Now, given two such functions
$F_1$ and $F_2$ associated to 
$f_1+\CI_0,f_2+\CI_0 \in\CC(\Gamma,M,C_0)$
and with gauge field $\epsilon_1$ and $\epsilon_2$ respectively,
one immediately computes the Poisson bracket 
$\{F_1,F_2\}_P=\Omega(\delta_{F_1},\delta_{F_2})$ and obtains
\begin{equation}
\{F_1,F_2\}_P=\Gamma^{ij}\epsilon_{1i}(0)\epsilon_{2j}(0)
\end{equation}
This coincides with the restriction to $C_0$ of
$\{f_1+\CI_0,f_2+\CI_0\}_{C_0}$ and defines a Poisson homomorphism
between $\CC(\Gamma,M,C_0)$ and the Poisson algebra of
$P(M,C_0,C_\pi)$. This homomorphism is $J_0^*$, the pull-back defined
by $J_0$, and the latter turns out to be a Poisson map. In an
analogous way we may show that $J_\pi$ is an anti-Poisson map and
besides
$$\{f_0\circ J_0, f_\pi\circ J_\pi\}=0\quad {\rm for\ any}\  
f_{u}\in\CC(\Gamma,M,C_{u}),\ {u}=0,\pi$$

These results can be
summarized in the following diagram
\begin{equation}
\begin{matrix}
&\displaystyle
J^*_0
&\displaystyle
&\displaystyle
J^*_\pi
&\displaystyle
\cr
\CC(\Gamma,M,C_0)
&\displaystyle
\longrightarrow
&\displaystyle
\CP
&\displaystyle
\longleftarrow
&\displaystyle
\CC(\Gamma,M,C_\pi)
\cr
\end{matrix}
\end{equation}
in which $J_0^*$ is a Poisson homomorphism, $J_\pi^*$ antihomomorphism
and the image of each map is the commutant (with respect to the
Poisson bracket) of the other. This can be considered as a
generalization of the symplectic dual pair to the context of Poisson
algebras.
 
\section{Quantization}

Since the work of Cattaneo and Felder \cite{CaFe} we know that the
perturbative expansion of certain Green's functions of the model
defined on the disc with free boundary conditions ($C=M$) yields the
deformation quantization of the Poisson bracket on $M$ proposed by
Kontsevich in ref. \cite{Kon}, namely
\begin{eqnarray}\label{Kontsevich}
<f(X(0))g(X(1))>_{X(\infty)=x}=f*g(x)
\end{eqnarray}
where $0$, $1$ and $\infty$
are three different (ordered) points at the boundary of the disc.

The Hamiltonian analysis and the topological nature of the model
suggest that for a general $C$ the perturbative expansion of
(\ref{Kontsevich}) should reproduce Kontsevich's formula for the reduced
Poisson bracket on the brane.

In a recent paper \cite{CaFe3} the same authors have computed the
Green's functions for coisotropic branes. Their result indicates that
under some technical assumptions about the gauge transformations at
the boundary one obtains again a deformation of the algebra of
functions invariant under the deformed gauge transformations. We want 
to stress that the gauge transformations at the boundary introduce technical
difficulties for the consistent quantization of the theory and make
the conclusion somewhat involved.

A different scenario in which one could have a cleaner derivation is
the case of constant rank Poisson-Dirac branes. Here there are not
gauge transformations acting at the boundary and the technical
requirements disappear. The natural guess is that the perturbative
quantization will give Kontsevich's formula for the Dirac bracket (see
\cite{CaFa} to check that the reduced Poisson bracket on a
Poisson-Dirac brane is the standard Dirac bracket).

At first sight things do not seem to work, as the propagator
corresponding to the perturbative expansion around the zero Poisson
structure does not exist with these BC. However, this is not very
surprising since to compute the Dirac bracket one has to invert the
matrix of Poisson brackets of constraints defining the constant rank
Poisson-Dirac brane. The appropriate expansion in this case is not
likely to be around the zero Poisson structure.

We are currently working on this problem and we can announce some
preliminary results. Consistent perturbative quantization of the
theory can be defined by changing the decomposition of the action into
the unperturbed part and the perturbation.  Proceeding in this way we
can show that at least in coordinates of the target adapted to the
brane such that the components of the Poisson structure are constant,
the perturbative quantization of the model produces the Kontsevich's
formula for the Dirac bracket on the brane. Details of the calculation
will appear elsewhere.

On the light of these facts it is natural to conjecture that the same
result on the deformation of the Dirac bracket (or some other
equivalent to it) holds for a general Poisson bracket on $M$ and a
general Poisson-Dirac brane. This will be the subject of further
research.

\noindent 
Iv\'an Calvo, Fernando Falceto \\
Depto. F\'{\i}sica Te\'orica, Univ. Zaragoza, \\
E-50009 Zaragoza, Spain\\
E-mail: {\tt icalvo@unizar.es, falceto@unizar.es}

\label{lastpage}
\end{document}